\documentclass[seceq,preprint]{ptptex}

\usepackage{epsfig, color,graphicx}
\usepackage{amsmath}
\usepackage{amssymb}
\newcommand{\nn}{\nonumber \\}
\newcommand{\bea}{\begin{eqnarray}}
\newcommand{\ena}{\end{eqnarray}}
\newcommand{\vs}[1]{\vspace{#1 mm}}
\newcommand{\hs}[1]{\hspace{#1 mm}}
\renewcommand{\a}{\alpha}

\renewcommand{\c}{\gamma}
\renewcommand{\d}{\delta}

\newcommand{\s}{\sigma}

\newcommand{\la}{\lambda}

\newcommand{\p}[1]{(\ref{#1})}
\newcommand{\tm}{\tilde{m}}
\newcommand{\tr}{\tilde{r}}

\newcommand{\squaret}{\kern1pt\vbox{\hrule height 0.9pt\hbox{\vrule width
0.9pt\hskip 2pt\vbox{\vskip 5.5pt}\hskip 3pt\vrule width 0.3pt}\hrule height
0.3pt}\kern1pt}
\begin{document}

\preprintnumber[3cm]{
KU-TP 024}
\title{\large  Black Holes in the Dilatonic Einstein-Gauss-Bonnet Theory
in Various Dimensions II \\
-- Asymptotically AdS Topological Black Holes --
}
\author{Zong-Kuan {\sc Guo},$^{a,b,}$\footnote{e-mail address: guozk at phys.kindai.ac.jp}
Nobuyoshi {\sc Ohta}$^{a,}$\footnote{e-mail address: ohtan at phys.kindai.ac.jp}
and Takashi {\sc Torii}$^{c,}$\footnote{e-mail address: torii at ge.oit.ac.jp}
}

\inst{
$^a$Department of Physics, Kinki University, Higashi-Osaka,
Osaka 577-8502, Japan\\
$^b$Fakult{\"a}t f{\"u}r Physik, Universit{\"a}t Bielefeld, Postfach 100131,
33501 Bielefeld, Germany\\
$^c$Department of General Education, Osaka Institute of Technology,
Asahi-ku, Osaka 535-8585, Japan
}
\abst{
We study asymptotically AdS topological black hole solutions with $k=0$
(plane symmetric)
in the Einstein gravity with Gauss-Bonnet term, the dilaton and a ``cosmological
constant'' in various dimensions.
We derive the field equations for suitable ansatz for general $D$ dimensions.
We determine the parameter regions including dilaton couplings where such solutions
exist and construct black hole solutions of various masses numerically in $D=4,5,6$
and 10 dimensional spacetime with $(D-2)$-dimensional hypersurface
of zero curvature.
}

\maketitle

\section{Introduction}

This is the second of a series of papers about the black hole
solutions in dilatonic Einstein-Gauss-Bonnet theory in higher dimensions.~\cite{GOT}

The primary motivation for the work is the following.
Many works have been done on black hole solutions in dilatonic gravity,
and various properties have been studied since the work in Refs.~\citen{GM}
and \citen{GHS}.
On the other hand, it is known that there are higher-order quantum corrections from
string theories.~\cite{GS} It is then natural to ask how these corrections may modify
the results. Several works have studied the effects of higher order
terms,~\cite{KMRTW,AP,TYM,CGO1,CGO2}
but most of the work done so far considers theories without
dilaton,~\cite{BD,GG,Cai1}
which is one of the most important ingredients in the string effective theories.
Hence it is important to study black hole solutions and their properties
in the theory with the higher order corrections and dilaton.
The simplest higher order correction is the Gauss-Bonnet (GB) term,
which may appear in heterotic string theories.

In our previous paper,~\cite{GOT} we have studied black hole solutions with
the GB correction term and dilaton for asymptotically flat solutions in various
dimensions from 4 to 10 with $(D-2)$-dimensional hypersurface of positive
curvature. A natural next problem is then to study such solutions with hypersurface
of zero curvature. It turns out that there do not exist solutions in such
theories, as we will discuss later. To construct such solutions, we find that it
is necessary to add a cosmological constant.
In the string perspective, it may be also more interesting to examine asymptotically
anti-de Sitter (AdS) black hole solutions with possible application to AdS/CFT
correspondence in mind.
In this paper, we present our results for asymptotically AdS solutions with
a negative cosmological constant. The class of solutions considered in
this paper with $(D-2)$-dimensional hypersurface of zero curvature
are known as topological black holes.

It may appear odd to add a cosmological
constant in a low-energy effective theory of the superstring theories,
but actually it may be present in such theories. For example, it is
known that type IIA theories have a 10-form whose expectation value
may give rise to such a cosmological constant.~\cite{POL} Other possible sources
include generation of such a term at one-loop in non-supersymmetric
heterotic string.~\cite{AGMV} There are also various forms in superstrings
which could produce similar terms with various dilaton
dependences, so we will simply suppose that such terms are present.

At this point, we should be careful about what we mean by a cosmological constant.
The above 10-form in type IIA theories gives a real cosmological
constant in the string frame. When transformed into the Einstein frame,
this gives rise to a term with dilaton coupling, i.e.,  a Liouville type
of potential for the dilaton. When we consider asymptotically AdS type behavior
of the metric, the dilaton coupling to the GB term produces another effective
potential of Liouville type. With only one of these terms, there would be
no asymptotically constant solution for dilaton, and the desirable
black hole solutions cannot be obtained.\footnote{
Black hole solutions in dilatonic Einstein-Maxwell theories with Liouville-type
potential but without GB term are studied in Refs.~\citen{PW,CHM}.
Exact solutions and their properties are discussed in Ref.~\citen{Ch}
in dilatonic Einstein theory with Liouville potential.
The presence of a Liouville potential already changes completely
the difficulty of the system.
In fact it is no longer an integrable system even in the absence of a GB term.
}
We find, however, that there are
interesting black hole solutions for suitable range of parameters of
these dilaton couplings.
We discuss the allowed parameter range where this potential together with GB
contribution can give asymptotically AdS black hole solutions.
For some choice of these couplings in the allowed region, we then construct
asymptotically AdS solutions in $D=4,5,6$ and 10 and discuss their properties.

This paper is organized as follows.
In \S~2, we first present the action of our consideration with GB and
cosmological terms, and give basic equations to solve. We then discuss
symmetry properties of the theory which will be useful in our following analysis.
In \S~3, we discuss the boundary conditions and asymptotic behaviors of
the black hole solutions and identify the allowed parameter range for the existence
of the black hole solutions by looking at the asymptotic expansion of
various fields. Here we also show that there is no black hole solution
if we do not have cosmological constant.
In \S~4, we first discuss how to generate black hole solutions with different
horizon radii and cosmological constants, given a solution for a certain
parameters. For comparison, we also summarize results for non-dilatonic
case.  We then present our black hole solutions in $D=4,5,6$ and 10 dimensions
for some typical choices of the allowed parameters, together with physical
quantities of the solutions. Using the scaling properties of the theory,
we determine the gravitational mass in terms of the cosmological constant
and horizon radius.
We conclude this paper with summary of our results and discussions of
remaining problems in \S~5.

\section{Dilatonic Einstein-Gauss-Bonnet theory}

\subsection{The action and basic equations}

We consider the following low-energy effective action for a
heterotic string
\bea
S=\frac{1}{2\kappa_D^2}\int d^Dx \sqrt{-g} \left[R - \frac12
 (\partial_\mu \phi)^2
 + \a_2 e^{-\c \phi} R^2_{\rm GB} -\Lambda e^{\la \phi} \right],
\label{act}
\ena
where $\kappa_D^2$ is a $D$-dimensional gravitational constant,
$\phi$ is a dilaton field, $\alpha_2=\a'/8$ is a numerical
coefficient given in terms of the Regge slope parameter $\a'$,
and
$R^2_{\rm GB} = R_{\mu\nu\rho\sigma} R^{\mu\nu\rho\sigma}
- 4 R_{\mu\nu} R^{\mu\nu} + R^2$ is the GB correction.
In this paper we leave the coupling constant of dilaton $\gamma$ arbitrary as much
as possible, while the ten-dimensional critical string theory predicts $\c=1/2$.
We have also included the negative cosmological constant $\Lambda=-(D-1)_2/\ell^2$
with possible dilaton coupling $\la$.
The RR 10-form in type IIA theory
can produce ``cosmological constant'' in the string frame, but that will
carry such dilaton couplings with $\la=\frac52$  in the Einstein frame.~\cite{POL}
Note that this ``cosmological term'' gives a Liouville type of potential.
If this is the only potential, there is no stationary point and the dilaton
cannot have a stable asymptotic value. However, for asymptotically
AdS solutions, the Gauss-Bonnet term produces an additional potential
in the asymptotic region, and we will see that it is possible to have
the solutions where the dilaton takes finite constant value at infinity.
There may be other possible sources of ``cosmological terms'' with different
dilaton couplings, so we leave $\lambda$ arbitrary and specify it
in the numerical analysis.

Varying the action~\p{act} with respect to $g_{\mu\nu}$, we obtain
the gravitational equation:
\begin{eqnarray}
&&
G_{\mu\nu}
-\frac12\biggl[\nabla_{\mu}\phi\nabla_{\nu} \phi -\frac12 g_{\mu\nu}(\nabla\phi)^2\biggr]
\nonumber \\
&&
~~~~~~~~~~~
+\alpha_2 e^{-\gamma\phi}\Bigl[H_{\mu\nu}
+4(\gamma^2\nabla^{\rho}\phi\nabla^{\sigma}\phi
-\gamma\nabla^{\rho}\nabla^{\sigma}\phi)P_{\mu\rho\nu\sigma}\Bigr]
+\frac12 g_{\mu\nu}\Lambda e^{\la\phi}
=0,
\label{GB-eq}
\end{eqnarray}
where
\begin{eqnarray}
&&G_{\mu\nu}\equiv R_{\mu\nu}-{1\over 2}g_{\mu\nu}R,
\\
&&H_{\mu\nu}\equiv 2\Bigl[RR_{\mu\nu}-2R_{\mu \rho}R^{\rho}_{~\nu}
-2R^{\rho\sigma}R_{\mu\rho\nu\sigma}
+R_{\mu}^{~\rho\sigma\lambda}R_{\nu\rho\sigma\lambda}\Bigr]
-{1\over 2}g_{\mu\nu}R^2_{\rm GB},
\\
&& P_{\mu\nu\rho\sigma}\equiv
R_{\mu\nu\rho\sigma}+2g_{\mu[\sigma}R_{\rho]\nu}
+2g_{\nu[\rho}R_{\sigma]\mu} +Rg_{\mu[\rho}g_{\sigma]\nu}.
\label{EGB:eq}
\end{eqnarray}
$P_{\mu\nu\rho\sigma}$ is the divergence free part of the Riemann tensor,
i.e.
\begin{equation}
\nabla_\mu P^{\mu}_{~\nu\rho\sigma}=0.
\end{equation}
The equation of the dilaton field is
\begin{eqnarray}
\label{dil-eq}
\squaret \phi -\alpha_2 \gamma e^{-\gamma\phi}  R^2_{\rm GB}
-\la \Lambda e^{\la\phi}=0,
\end{eqnarray}
where $\squaret$ is the $D$-dimensional d'Alembertian.

We parametrize the metric as
\bea
ds_D^2 = - B e^{-2\d} dt^2 + B^{-1} dr^2 + r^2 h_{ij}dx^i dx^j,
\ena
where $h_{ij}dx^i dx^j$ represents the line element of a
$(D-2)$-dimensional hypersurface with constant curvature
$(D-2)(D-3)k$ and volume $\Sigma_k$ for $k=\pm 1,0$.
We consider the plane symmetric case $k=0$ for the black hole solutions in this paper.

The metric function $B=B(r)$ and the lapse function $\d=\d(r)$ depend only on the
radial coordinate $r$. The field equations can be read off from Ref.~\citen{GOT}
or \citen{BGO} as
\bea
&& \bigl[(k-B)\tr^{D-3}\bigr]' \frac{D-2}{\tr^{D-4}}h -\frac12 B \tr^2 {\phi'}^2
 - (D-1)_4\,e^{-\c\phi}\frac{(k-B)^2}{\tr^2} \nn
&& \hs{10} + 4(D-2)_3\, \c e^{-\c\phi}B(k-B)(\phi''-\c {\phi'}^2) \nn
&& \hs{10} + 2(D-2)_3\,\c e^{-\c\phi}\phi'\frac{(k-B)[(D-3)k-(D-1)B]}{\tr}
-\tr^2 \tilde \Lambda e^{\la \phi}= 0\,,
\label{f1} \\
&& \delta'(D-2)\tr h + \frac12 \tr^2 {\phi'}^2
 -2(D-2)_3\, \c e^{-\c\phi}(k-B)(\phi''-\c {\phi'}^2) =0 \,,
 \label{f2} \\
&&
(e^{-\d} \tr^{D-2} B \phi')' = \c (D-2)_3 e^{-\c\phi-\d} \tr^{D-4}
\Big[ (D-4)_5 \frac{(k-B)^2}{\tr^2} + 2(B'-2\d' B)B' \nn
&& \hs{10} -4(k-B)BU(r)
-4\frac{D-4}{\tr}(B'-\d'B)(k-B) \Big] + e^{-\d} \tr^{D-2} \la \tilde\Lambda e^{\la\phi},
\label{f3}
\ena
where we have defined the dimensionless variables: $\tr \equiv r/\sqrt{\a_2}$,
$\tilde \Lambda = \a_2 \Lambda$, and the primes in the field equations
denote the derivatives with respect to $\tr$. Namely we measure our length
in the unit of $\sqrt{\a_2}$. We have kept $k$ in these equations and defined
\bea
(D-m)_n &\equiv& (D-m)(D-m-1)(D-m-2)\cdots(D-n), \nn
h &\equiv& 1+2(D-3) e^{-\c\phi} \Big[ (D-4) \frac{k-B}{\tr^2}
 + \c \phi'\frac{3B-k}{\tr}\Big], \\
\tilde h &\equiv& 1+2(D-3) e^{-\c\phi} \Big[(D-4)\frac{k-B}{\tr^2}
+\c\phi'\frac{2B}{\tr} \Big], \\
U(r) &\equiv& (2 \tilde h)^{-1} \Bigg[ (D-3)_4 \frac{k-B}{\tr^2 B}
-2\frac{D-3}{\tr}\Big(\frac{B'}{B}-\d'\Big) -\frac12 \phi'^2 \nn
&& + (D-3)e^{-\c\phi} \Bigg\{ (D-4)_6 \frac{(k-B)^2}{\tr^4 B}
- 4 (D-4)_5 \frac{k-B}{\tr^3}\Big(\frac{B'}{B}-\d'-\c\phi'\Big) \nn
&& -4(D-4)\c \frac{k-B}{\tr^2}\Big( \c \phi'^2 +\frac{D-2}{\tr}\phi'-\Phi \Big)
+8 \frac{\c\phi'}{\tr} \biggl[\Big(\frac{B'}{2}-\d' B\Big)\Big(\c\phi'-\d'
+\frac{2}{\tr} \Big) \nn
&& -\frac{D-4}{2\tr}B' \biggr] + 4(D-4)\Big(\frac{B'}{2B}-\d' \Big)
\frac{B'}{\tr^2}-\frac{4\c}{\tr}\Phi (B'-2\d'B)\Bigg\}
-\frac{1}{B} \tilde \Lambda e^{\la \phi}\Biggr],\\
\Phi &\equiv& \phi'' +\Big(\frac{B'}{B}-\d' +\frac{D-2}{\tr}\Big) \phi'.
\label{dil}
\ena

\subsection{symmetry and scaling}

It is useful to consider several symmetries of our field equations
(or our model).
Firstly the field equations are invariant under the transformation:
\bea
\gamma \to -\gamma, ~~
\lambda \to -\lambda, ~~
\phi \to -\phi\, .
\label{sym0}
\ena
By this symmetry, we can restrict the parameter range of $\c$ to $\c\geq 0$.

For $k=0$, the field equations~\p{f1}--\p{f3} are
invariant under the scaling transformation
\bea
B \to a^2 B, ~~
\tr \to a \tr,
\label{sym1}
\ena
with an arbitrary constant $a$.
If a black hole solution with the horizon radius $\tr_H$ is obtained,
we can generate solutions with different horizon radii but the same $\tilde \Lambda$
by this scaling transformation.

The field equations~\p{f1}--\p{f3} have a shift symmetry:
\bea
\phi \to \phi-\phi_{\ast}, ~~
\tilde\Lambda \to e^{(\la-\c)\phi_{\ast}} \tilde\Lambda, ~~
B \to e^{-\c \phi_{\ast}} B\, ,
\label{sym2}
\ena
where $\phi_{\ast}$ is an arbitrary constant.
This changes the magnitude of the cosmological constant. Hence this may
be used to generate solutions for different cosmological constants but
with the same horizon radius, given a solution for some cosmological constant
and $\tr_H$.

The final one is another shift symmetry under
\bea
\delta \to \delta - \delta_{\ast}, ~~
t \to  e^{-\delta_{\ast}}t,
\label{sym3}
\ena
with an arbitrary constant $\delta_{\ast}$, which may be used to shift
the asymptotic value of $\delta$ to zero.

The model \p{act} has several parameters $D$, $\alpha_2$, $\Lambda$, $\gamma$,
and $\lambda$. The black hole solutions have also physical independent
parameters such as the horizon radius $\tr_H$ and the value of $\delta$ at infinity.
However, owing to the above symmetries (including the scaling by $\alpha_2$),
we can reduce the number of the parameters and are left only with
$D$, $\gamma\geq 0$, $\lambda$, and $\tr_H$.

\section{Boundary conditions and asymptotic behavior}

We study plane symmetric solution with $k=0$ and a negative cosmological
constant $\tilde \Lambda <0$. In this section, we discuss the boundary conditions
and asymptotic behaviors of the metric and the dilaton fields. In this process,
we will see that there is no black hole solution for $k=0$ without the cosmological
constant.

\subsection{Regular horizon}

Let us first examine the boundary conditions of the black hole spacetime.
We assume the following boundary conditions for the metric functions:
\begin{enumerate}
\item
The existence of a regular horizon $\tr_H$:
\bea
\label{hor}
B(\tr_H)=0, ~~
|\d_H| < \infty, ~~
|\phi_H|< \infty\, .
\ena
\item
The nonexistence of singularities outside the event horizon ($\tr > \tr_H$):
\bea
B(\tr)>0, ~~
|\d| < \infty, ~~
|\phi|< \infty\, .
\ena
\end{enumerate}
Here and in what follows, the values of various quantities at the horizon are
denoted with subscript $H$.
At the horizon, it follows from~\p{f1}--\p{dil} that
\bea
&& B_H=0, ~~
h_H=\tilde h_H=1,\nn
&& B_{H}'= - \frac{\tilde\Lambda}{D-2} \tr_H e^{\la\phi_H}, \nn
&& \phi_{H}'= -\frac{1}{\tr_H}\Bigl[2\c(D-3)\tilde\Lambda e^{(\la-\c)\phi_H}
+ (D-2)\la \Bigr], \nn
&& \d_{H}'= - \frac{1}{2(D-2)} \tr_H(\phi_H')^2.
\label{bhor}
\ena

{}From these equations, we see that all the derivatives of these quantities
vanish at the horizon for $\tilde \Lambda=0$,\footnote{
When the cosmological constant is zero, $\la$ is absent.}
and our basic equations~\p{f1}--\p{f3}
tells us that these fields are constant, giving no nontrivial solutions.
This is the basic reason why we consider these topological solutions
with cosmological constant.

\subsection{Asymptotic behavior at infinity and the effective potential}

At infinity we assume the condition that the leading term of the metric function
$B$ comes from AdS radius $\tilde{\ell}_{AdS}$, i.e.,
\begin{enumerate}
\item[3.]
``AdS asymptotic behavior" ($\tr \to \infty$):
\bea
\label{as}
B \sim \tilde{b}_2 \tr^2 - \frac{2\tilde M}{\tr^{\mu}}, ~~~
\d(r) \sim \d_0 + \frac{\d_1}{\tr^{\s}}, ~~
\phi \sim \phi_0 + \frac{\phi_1}{\tr^{\nu}} \,,
\ena
with finite constants $\tilde{b}_2>0$, $\tilde M$, $\d_0$, $\d_1$,  $\phi_0$, $\phi_1$
and positive constant $\mu$, $\s$, $\nu$.
\end{enumerate}
The coefficient of the first term $\tilde{b}_2$ is related to the AdS radius
as $\tilde{b}_2 = \ell_{\rm AdS}^{-2}$. However, this condition is not sufficient
for the spacetime to be the exactly AdS asymptotically. Strictly
speaking, the asymptotically AdS spacetime is left invariant under
$SO(D-1,2)$.~\cite{Henneaux} Whether the solution satisfies the AdS-invariant
boundary condition or not depends on the value of the power indices
$\mu$, $\s$, and $\nu$.

If $\phi \to -\infty$ and curvature tensors of the spacetime becomes small
enough asymptotically at infinity, the model can be well approximated
by Einstein gravity with a single scalar field with potential.
However, we will not consider such solution in this paper but briefly comment
on such possibility in \S~\ref{CD}.

Let us now briefly analyze the effective potential picture which is helpful
to understand the asymptotic behaviors of our dilatonic system.
We write the equation of the dilaton field as
\begin{eqnarray}
\tilde \squaret \phi -\frac{d\tilde{V}_{\rm eff}}{d\phi}=0,
\label{dil_eff}
\end{eqnarray}
where the ``effective potential'' is defined by
\begin{eqnarray}
\label{effective-potential}
\tilde{V}_{\rm eff}= - e^{-\gamma\phi}\tilde R^2_{\rm GB}+\tilde \Lambda e^{\lambda\phi}.
\end{eqnarray}
Here the tilde over GB term means that it is evaluated using $\tr$.
The constant $\la$ is determined by the way how the cosmological constant
is introduced.\footnote{
When the stiff (or pure) cosmological constant is introduced in the string frame
in $D$ dimensions, $\lambda$ in our Einstein frame is $\frac{D}{\sqrt{2(D-2)}}$.}
The field equation~(\ref{dil_eff}) is written as
\begin{eqnarray}
\label{dil-infty}
\frac{1}{e^{-\delta}\tr^{D-2}}\Big(e^{-\delta}\tr^{D-2}B\phi'\Bigr)'
=\frac{d\tilde{V}_{\rm eff}}{d\phi},
\end{eqnarray}
and it is pointed out that the dilaton field climbs up the potential slope.~\cite{TMN}
(Note that the sign of the r.h.s. is opposite to the homogeneous and
time-dependent case where the dilaton field rolls down the potential slope.)

For the asymptotic behavior for $B$ in Eq.~\p{as},
this potential reduces asymptotically to
\begin{eqnarray}
\label{pot_inf}
\tilde{V}_{\rm eff}= - (D)_3\; \tilde{b}_2^{\:2}\; e^{-\gamma\phi}
+ \tilde \Lambda e^{\lambda\phi}.
\end{eqnarray}
When $\lambda =0$, the cosmological term decouples from the dilaton field
but minimally couples through gravity. The effective potential becomes, up to a constant,
\begin{eqnarray}
\tilde{V}_{\rm eff}=  -(D)_3\; \tilde{b}_2^2\; e^{-\gamma\phi},
\end{eqnarray}
and the dilaton field climbs up the potential and diverges for $\phi \to +\infty$.
(Remember that $\gamma>0$.) From the asymptotic expansion, we find that
the dilaton field behaves as $\phi \sim \frac{1}{\gamma} \log(\log r)$,
and breaks the asymptotic AdS-invariant condition.
This will be confirmed in the next subsection.

\begin{figure}[ht]
\begin{center}
\includegraphics[width=6cm]{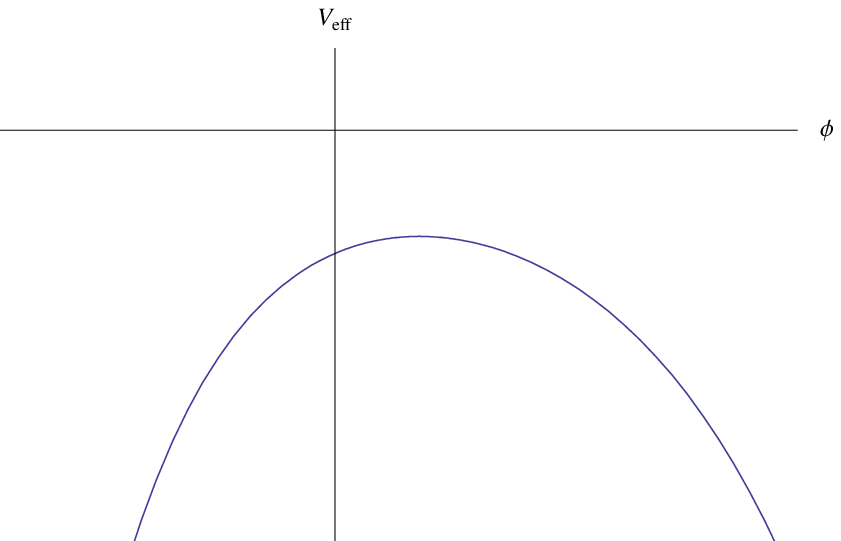}
\put(-110,-20){(a)}
\put(125,-20){(b)}
\hs{20}
\includegraphics[width=6cm]{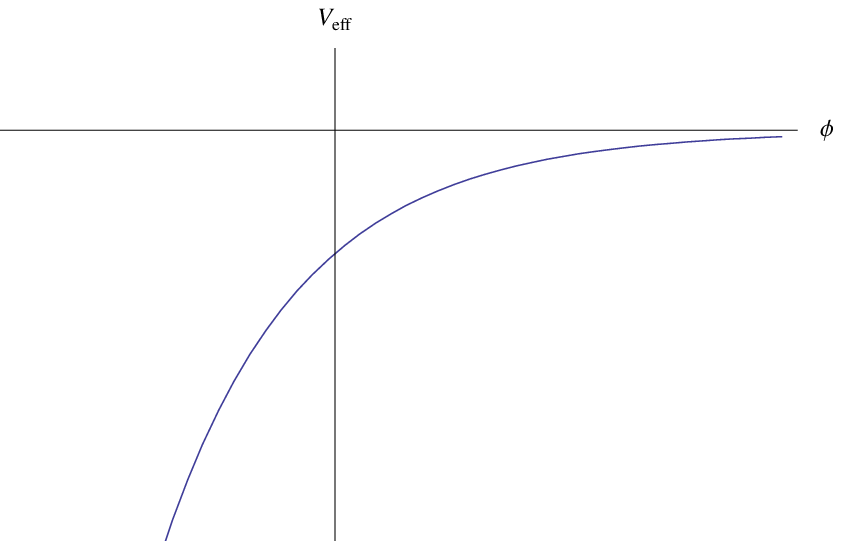}
\end{center}
\caption{The effective potentials of the dilaton field
in the Liouville potential case with (a) $\lambda >0$  and (b) $\lambda <0$.
}
\label{potential}
\end{figure}

When $\lambda>0$,  the effective potential (\ref{pot_inf})
has a maximum (Fig.~\ref{potential} (a)),
and the dilaton field would approach a finite constant $\phi_{0}$ at $r=\infty$.
Thus at infinity, the dilaton field should stay at the maximum of the potential,
and it is expected that the spacetime is ordinary AdS asymptotically.
For $\la<0$, the effective potential monotonically increases (Fig.~\ref{potential} (b)),
and cannot give AdS-invariant spacetime. We do not consider this case
and concentrate on $\la>0$.

It should be noted, however,  that the ``effective potential" is not the ordinary one
since it contains metric functions explicitly which depend on $\tilde{r}$.
The configuration of the ``effective potential" changes depending on
$\tilde{r}$, and there is a case where it does not give a right asymptotic
behavior, e.g., the dilaton field diverges although the form of the potential
is of the type in Fig.~\ref{potential}\:(a).
Hence we analyse the asymptotic behaviors of the field functions in detail
by looking at the asymptotic expansion in the following.

\subsection{Asymptotic expansion}

Substituting Eqs.~(\ref{as}) into the field equations (\ref{f1}) and
(\ref{f3}), one finds the conditions that the leading terms ($\tr^2$ and constant
terms in each equation) balance with each other are given by
\bea
&& \label{phi_inf_1}
(D)_3\gamma\; e^{-\gamma\phi_0} \tilde{b}_2^{~2}
+ \lambda \tilde{\Lambda}e^{\lambda\phi_0} = 0 \,, \\
&& \label{B_inf_1}
(D-1)_4\;e^{-\gamma\phi_0} \tilde{b}_2^{~2}-(D-1)_2 \tilde{b}_2
-\tilde{\Lambda}e^{\lambda\phi_0} = 0 \,,
\ena
which determine $\tilde b_2$ and $\phi_0$,
while $\d_0$ can be arbitrary because only its derivative appears in our field
equations.
Since $\gamma$ is positive and $\tilde \Lambda$ is negative, $\lambda$ should
be also positive by Eq.~\p{phi_inf_1}. This restricts the parameter space
to $\lambda>0$. From these equations, we find
\bea
\label{inf_1}
(D)_2 \gamma \tilde{b}_2   +\bigl[(D-4)\lambda
+D\gamma\bigr]\tilde\Lambda e^{\lambda \phi_{0}}=0,\\
\label{inf_2}
(D-3)\bigl[(D-4)\lambda +D\gamma\bigr]\tilde{b}_2 -\lambda e^{\gamma \phi_{0}}=0.
\ena

For $\lambda = \gamma$, the cosmological constant and $\phi_0$ are found to be
\begin{equation}
\label{gamma=lambda}
\tilde{\Lambda}=-\frac{(D)_1}{4(D-2)_3}, ~~~~~
e^{-\gamma \phi_{0}}\tilde{b}_2=\frac{1}{2(D-2)_3}.
\end{equation}
If we allow the possibility that the dilaton diverges whereas $\tilde b_2$
becomes infinity, there may be other solutions but the expansion does not
give sensible result for such a case.
Note that due to the shift symmetry (\ref{sym2}), the values of $\phi_0$ and
$\tilde b_2$ themselves are not determined individually.

For $\lambda\ne \gamma$,
Eqs.~(\ref{phi_inf_1}) and (\ref{B_inf_1}) give
\begin{eqnarray}
&&
\label{b2_inf}
\tilde{b}_2^{~2}
=\frac{-\lambda\tilde{\Lambda}}{(D)_3 \gamma}
\biggl[\frac{D(D-3)}{(D-1)_2}(-\tilde{\Lambda})\frac{\gamma}{\lambda}
\biggl(1+\frac{(D-4)\lambda}{D\gamma}\biggr)^2
\biggr]^{\frac{\gamma+\lambda}{\gamma-\lambda}},
\\
&&
\label{phi_inf_2}
e^{\phi_0}
=\biggl[\frac{D(D-3)}{(D-1)_2}(-\tilde{\Lambda})\frac{\gamma}{\lambda}
\biggl(1+\frac{(D-4)\lambda}{D\gamma}\biggr)^2
\biggr]^{\frac{1}{\gamma-\lambda}}.
\end{eqnarray}

The candidates of the next leading terms for Eqs.~\p{f1}--\p{f3} are
respectively given by
\begin{eqnarray}
&&
2(D-2)\left[\mu-(D-3)\right]\left[1-2(D-3)_4 \tilde{b}_2e^{-\c\phi_0}\right]\tilde{M}
\tr^{-\mu} \nonumber \\
&& \hs{10} - \Bigl[(D-2)_3\c \tilde{b}_2^2e^{-\c\phi_0}
\{4\nu^2-4(D-2)\nu+(D-1)(D-4)\} + \la \tilde{\Lambda} e^{\la\phi_0}
\Bigr]\phi_1 \tr^{2-\nu} \,,
\label{next1} \\
&& \label{next2} \rule[0mm]{0mm}{7mm}
\Bigl[2(D-3)_4 \tilde{b}_2 e^{-\c\phi_0}-1\Bigr]\sigma \d_1 \tr^{-\sigma}
 + 2(D-3) \tilde{b}_2 e^{-\c\phi_0}\gamma\nu(1+\nu)\phi_1 \tr^{-\nu} \,,\\
&& \rule[0mm]{0mm}{7mm}
4(D-2)_3\tilde{b}_2^2e^{-\c\phi_0}\c \sigma (\sigma-D)\d_1 \tr^{-\sigma}
 + \Bigl[D_3\tilde{b}_2^2e^{-\c\phi_0}\c^2 - \la^2 \tilde{\Lambda} e^{\la\phi_0}
 - \tilde{b}_2(D-1)\nu + \tilde{b}_2 \nu^2 \Bigr]\phi_1 \tr^{-\nu} \nonumber \\
&& \hs{10} - 4(D-2)_3\tilde{b}_2e^{-\c\phi_0}[\mu-(D-2)][\mu-(D-3)]\c \tilde{M}
\tr^{-\mu-2}
\label{next3}\,,
\end{eqnarray}
which should vanish.
By use of the leading equations (\ref{inf_1}) and (\ref{inf_2}), these equations
reduce to
\begin{eqnarray}
&& \label{next1-2}
\bigl[\mu-(D-3)\bigr]\bigl[(D-4)\lambda -D\gamma\bigr]\tilde{M}\tr^{-\mu}
- 2 \gamma\lambda(\nu+1)\bigl[\nu-(D-1)\bigr]\tilde{b}_2\phi_1\tr^{2-\nu}
\,,\\
&& \label{next2-2}
\rule[0mm]{0mm}{7mm}
\bigl[(D-4)\lambda-D\gamma\bigr]\sigma \d_1 \tr^{-\sigma}
+ 2\lambda \gamma\nu(\nu+1)\phi_1 \tr^{-\nu} \,,\\
&&
\rule[0mm]{0mm}{7mm}
4(D-2)\lambda\gamma\sigma(\sigma-D)\tilde{b}_2\delta_1 \tr^{-\sigma}
\nonumber \\
&& \hs{10}
+\Bigl\{(D)_2\lambda\gamma(\lambda+\gamma)
+\bigl[(D-4)\lambda+D\gamma\bigr]\nu\bigl[\nu-(D-1)\bigr]\Bigr\}\tilde{b}_2\phi_1
\tr^{-\nu}
\nonumber \\
&& \hs{10}
+ 4(D-2)\lambda\gamma\bigl[\mu-(D-2)\bigr]\bigl[\mu-(D-3)\bigr]\tilde{M}\tr^{-\mu-2}
\,,
\label{next3-2}
\end{eqnarray}
up to overall factors. Now we need to discuss two cases separately.

\subsubsection{$(D-4)\lambda-D\gamma \ne 0$ case}

Let us first consider the case with $(D-4)\lambda-D\gamma\ne 0$.
There are two different classes which give consistent expansions.
One is realized when the $\tr^{-\mu}$ term dominates over other terms.
We then find $\mu=D-3$ and $\nu>D-1$ and rename the coefficient $\tilde M$
as $\tilde M_0$.

The other class corresponds to the ordinary modes of the second order
differential equation of the dilaton field, where all these terms are
of the same order with $\mu=\nu-2=\sigma-2$.\footnote{
This behavior is different from the case of minimally coupled scalar field
with potential in general relativity, where the Breitenlohner and Freedmann
bound is discussed. There, asymptotic expansion gives
the relation $\mu=2\nu-2$.
}
{}From the next leading terms in (\ref{next1-2}) and  (\ref{next2-2}), we find
\bea
&&
\bigl[(D-4)\lambda -D\gamma\bigr]\tilde{M}
- 2 \lambda\gamma(\nu+1)\tilde{b}_2\phi_1=0
\,,\nn
&&
\rule[0mm]{0mm}{7mm}
\bigl[(D-4)\lambda-D\gamma\bigr] \d_1
 + 2\lambda \gamma(\nu+1)\phi_1 =0\,.
\label{nextcond}
\ena
Substituting these into the condition obtained from (\ref{next3-2})
\bea
&& 4 (D-2) (\nu-D) \la \nu \tilde b_2\d_1 + (D)_2 \la\c(\la+\c) \tilde b_2\phi_1 \nn
&&~~ + \nu (\nu+1-D)\bigl[(D-4)\la+D\c \bigr]\tilde b_2\phi_1
+ 4 (D-2)(\nu-D)(\nu+1-D)\la \c \tilde M=0,~~~~
\ena
we find
\bea
\nu =\nu_{\pm} = \frac{D-1}2 \left[1 \pm \sqrt{1 - \frac{4(D)_2\la\c(\la-\c)
\bigl[(D-4)\la+D\c\bigr]}{(D-1)^2\bigl[(D-4)^2\la^2-D^2\c^2-8(D-1)_2\la^2\c^2\bigr]}}
\;\right].
\label{nu}
\ena
The power indices $\nu$, $\mu$ and $\sigma$ do not depend on $\tilde\Lambda$.
Here we assume
\bea
(D-4)^2\la^2-D^2\c^2-8(D-1)_2\la^2\c^2\ne 0,
\ena
since otherwise $\nu$ has no solution.

We rewrite the indices $\nu_{\pm}$ as
\begin{eqnarray}
\nu_{\pm}= \frac{D-1}{2}\biggl[1\pm \sqrt{1-\frac{{\tilde{m}}^2}{{\tilde{m}}_{BF}^2}}
\biggr],
\label{nu-pm}
\end{eqnarray}
where the mass square of Breitenlohner and Freedman (BF) bound is defined by~\cite{BF}
\begin{eqnarray}
\tilde{m}_{BF}^2&&=-\frac{(D-1)^2}{4\tilde{\ell}_{AdS}^2}
=-\frac{(D-1)^2}{4}\tilde{b}_2,
\end{eqnarray}
and we define the mass square of the dilaton field as
\begin{eqnarray}
\tilde{m}^2 = - \frac{(D)_2\la\c(\la-\c)\bigl[(D-4)\la+D\c\bigr]}
 {(D-4)^2\la^2-D^2\c^2-8(D-1)_2\la^2\c^2}\tilde{b}_2.
\end{eqnarray}
by the analogy with the discussion in BF bound. This mass is considered to be
the second derivative of the potential of the dilaton field where
the $\tr$-dependence of the ``effective potential"~(\ref{effective-potential})
is taken into account.
Not that these equations hold even for the $\gamma=\lambda$ case if the value
of the cosmological constant is given by Eq.~(\ref{gamma=lambda}),
and $\nu_{\pm}=0,\; D-1$.

Let us now consider the normalizability of the dilaton field.
In the ordinary discussion of the BF bound, the normalizable condition \cite{BF,BKL,HM}
is assumed to be $\nu \geq (D-3)/2$.
For $\tilde{m}^2\geq\tilde{m}_{BF}^2+\tilde{\ell}_{AdS}^{-2}$,
we find that the $\nu_{-}$ mode is non-normalizable while the $\nu_{+}$ mode
is normalizable. Hence the $\nu_{-}$ mode should be tuned to vanish.
For $\tilde{m}_{BF}^2<\tilde{m}^2<\tilde{m}_{BF}^2 +\tilde{\ell}_{AdS}^{-2}$,
both modes are normalizable, and the spacetime has different classes of
AdS spacetime asymptotically depending on the ratio of these modes.
This normalizability condition does not seem to apply to our case
because our dilaton field is not considered to be quantum fluctuations.
Nevertheless, we adopt the boundary condition that the $\nu_{-}$ mode vanishes.
In the BF bound analysis, it is known that the normalizable modes give
finite conserved mass.
Although our system is different from such a system due to the GB term
and the dilaton coupling, we impose this condition.\footnote{
Although we do not prove if the mode with $\nu_+$ really gives the finite conserved mass
in this paper, it is expected that it is the case from the comparison of
asymptotic dependences of our model and those in general relativity with
the scalar field. This is under investigation, and we will discuss the problem
in the near future.
}

We eliminate the $\nu_{-}$ mode by tuning the value of $\phi_H$.
Hence $\phi_H$ is not a free parameter but a kind of shooting parameter
and should be chosen suitably for each horizon radius and
other theoretical parameters.
Also using the symmetry~\p{sym3}, we set $\d_0=0$.

The asymptotic forms of the field functions are then
\begin{eqnarray}
&&
\phi \sim \phi_0
+ \frac{\phi_{+}}{\tr^{\nu_{+}}}
+ \cdots\,,\nn
&&
\label{behavior}
B \sim \tilde{b}_2 \tilde{r}^2 
- \frac{2 \tilde{M}_{+}}{\tr^{\nu_{+}-2}} - \frac{2\tilde{M}_0}{\tr^{D-3}} + \cdots\,,\\
&&
\delta \sim \delta_0
+ \frac{\delta_{+}}{\tr^{\nu_{+}}}+ \cdots \, .\nonumber
\end{eqnarray}
Note that while $B$ has the term $r^{-\nu_{+}+2}$,
the $g_{tt}$ component of the metric behaves as
\begin{eqnarray}
-g_{tt}=Be^{-2\delta} \sim \tilde{b}_2 \tilde{r}^2
- \frac{2\tilde{M}_0}{\tr^{D-3}} + \cdots.
\end{eqnarray}
(Remember that we have chosen $\d_0=0$.)
This value of $\tilde{M}_0$ is the gravitational mass of the black holes.
Thus it is convenient to define the mass function $\tm_g(\tr)$ by
\bea
-g_{tt} = \tilde b_2 \tr^2 - \frac{2\tm_g(\tr)}{\tr^{D-3}}.
\label{mass}
\ena
We will present our results in terms of this function.

\subsubsection{$(D-4)\lambda-D\gamma= 0$ case}

In this case, from Eqs.~\p{nextcond}, we find $\nu=-1$.
This is a growing mode which means that the expansion is not valid.
We have then examined next order equations, but did not obtain any other
non-growing modes. Thus this case does not seem to give sensible solutions.
Hence we do not consider this case.




\subsection{Allowed parameter regions}

In this subsection, using the above results, we discuss the parameter regions
which give desirable black hole solutions.

\begin{figure}[th]
\hspace*{-2mm}
\includegraphics[width=3.8cm]{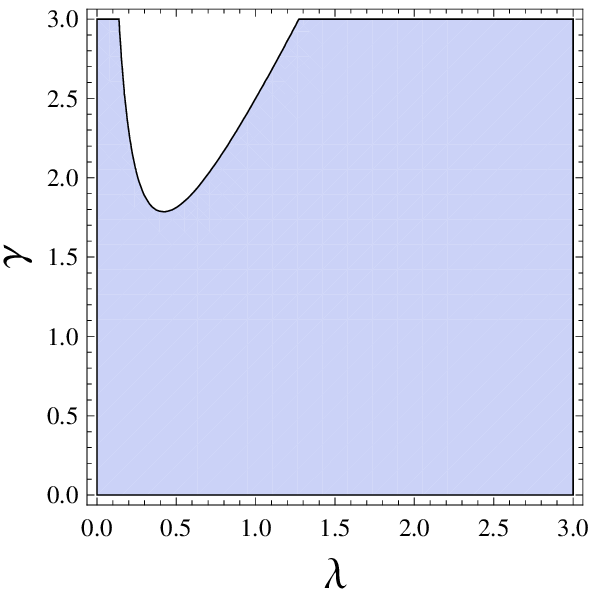}~
\includegraphics[width=3.8cm]{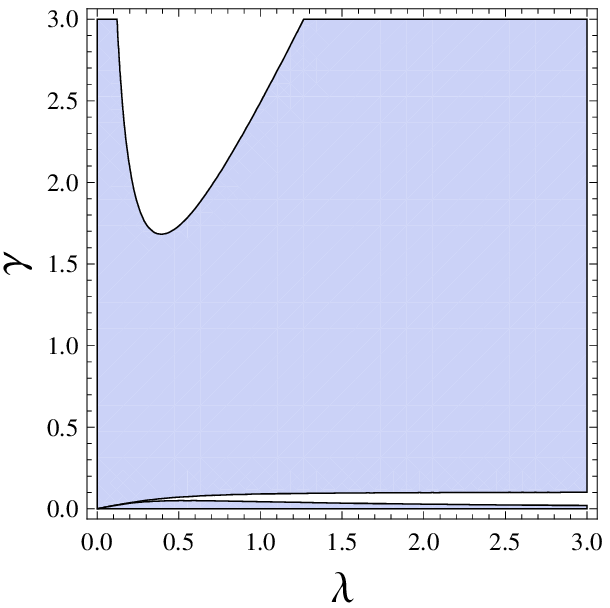}~
\includegraphics[width=3.8cm]{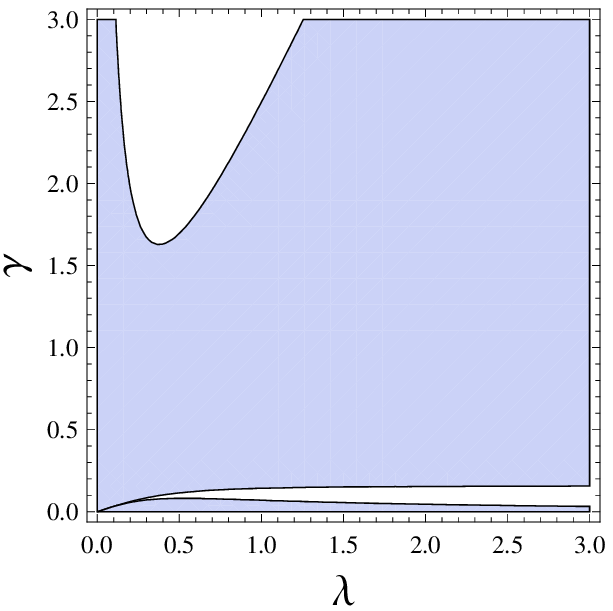}~
\includegraphics[width=3.8cm]{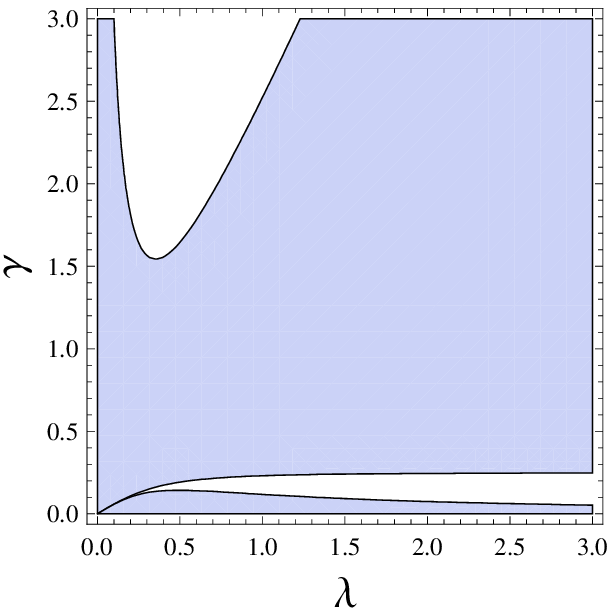}\\
\vspace{5mm}
\put(50,0){(a)}
\put(166,0){(b)}
\put(283,0){(c)}
\put(398,0){(d)}
\caption{The allowed regions for the parameters $\gamma$ and $\lambda$ from
the condition~\p{condition_1} for
(a) $D=4$, (b) $D=5$, (c) $D=6$, and (d) $D=10$.
}
\label{parameter-region_1}
\end{figure}

The mass of the dilaton field $\tilde{m}$ should satisfy the conditions
\begin{equation}
\tilde{m}_{BF}^2\leq \tilde{m}^2,
\label{BFbound}
\end{equation}
which comes from that the stability of the asymptotic structure of
the solution against time-dependent perturbations. This implies that
the parameters $\gamma$ and $\lambda$ should satisfy
\begin{equation}
\frac{4(D)_2\la\c(\la-\c)\bigl[(D-4)\la+D\c\bigr]}
{(D-1)^2\bigl[(D-4)^2\la^2-D^2\c^2-8(D-1)_2\la^2\c^2\bigr]}\leq 1.
\label{condition_1}
\end{equation}
The regions in which these conditions~(\ref{condition_1}) are satisfied
is depicted in Fig.~\ref{parameter-region_1}. Remember that we are considering
only the region $\c>0$, $\la>0$.

\begin{figure}[th]
\hspace*{-2mm}
\includegraphics[width=3.8cm]{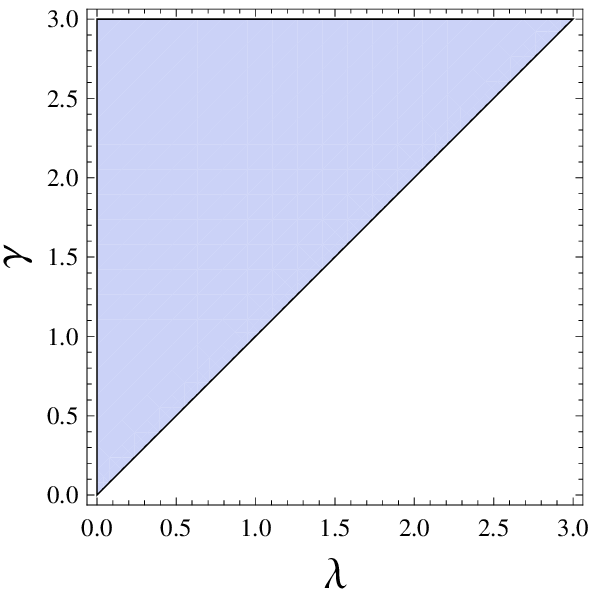}~
\includegraphics[width=3.8cm]{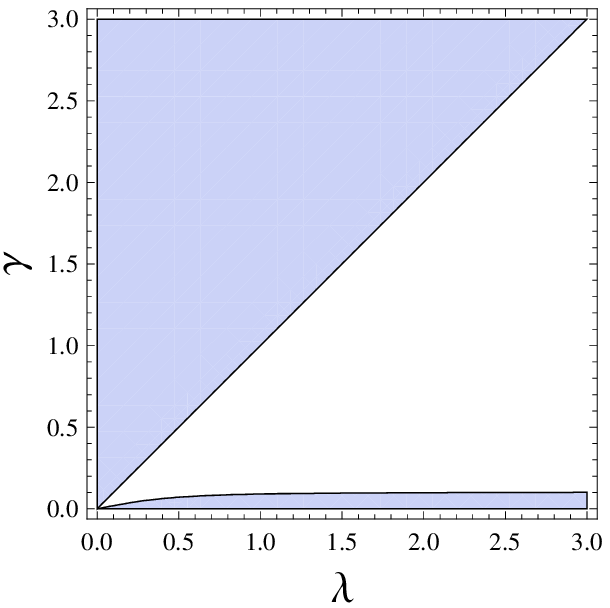}~
\includegraphics[width=3.8cm]{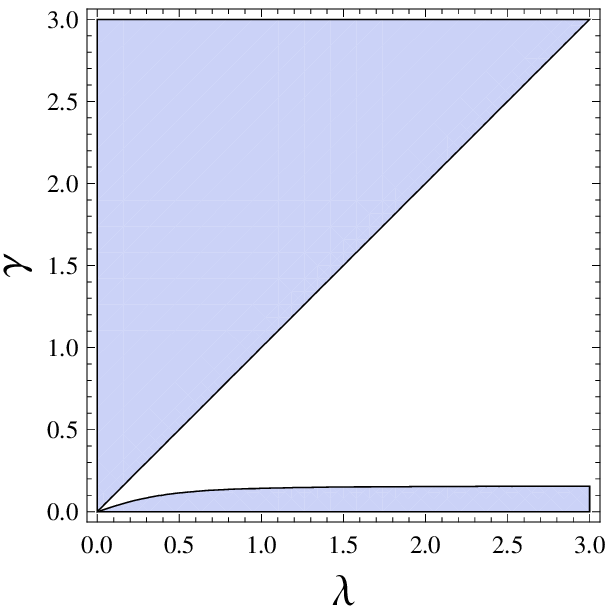}~
\includegraphics[width=3.8cm]{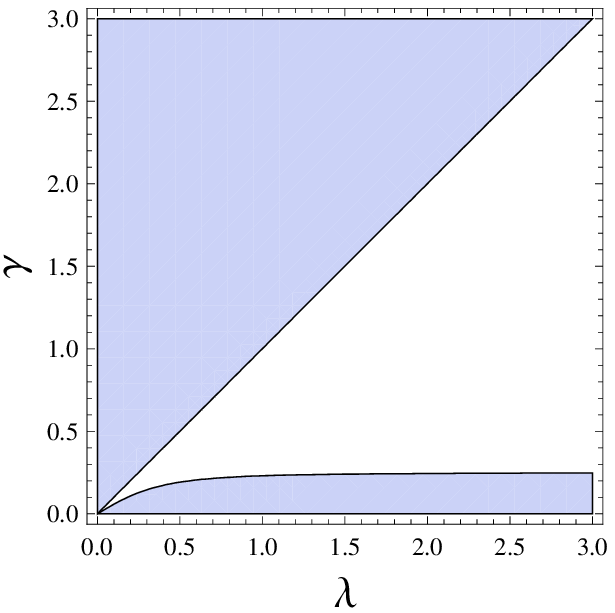}\\
\vspace{5mm}
\put(50,0){(a)}
\put(166,0){(b)}
\put(283,0){(c)}
\put(398,0){(d)}
\caption{The allowed regions for the parameters $\gamma$ and $\lambda$ from
the condition~\p{condition_finite-dilaton} for
(a) $D=4$, (b) $D=5$, (c) $D=6$, and (d) $D=10$.
}
\label{parameter-region_2}
\end{figure}

We also impose another condition
\begin{equation}
\tilde{m}^2<0.
\label{condition_finite-dilaton}
\end{equation}
When $\tilde{m}^2>0$, the potential of the dilaton field is an ordinary
(non-tachyonic) convex potential, and there is a growing mode according to
Eq.~(\ref{nu-pm}). Such a mode should be eliminated by tuning $\phi_H$ at
the horizon for the finiteness
of the dilaton field. The solution must be stable against time-dependent
perturbations since the condition of BF bound (\ref{BFbound}) is satisfied.
By the numerical analysis, however, we find that the growing mode cannot be
eliminated just by tuning $\phi_H$, and the dilaton field diverges.
Although this fact does not mean that there cannot be a solution with $\tilde{m}^2<0$
for any parameters $\gamma$ and $\lambda$, we impose the
condition~(\ref{condition_finite-dilaton}) in this paper.
Then the potential of the dilaton field is tachyonic, and
the dilaton field climbs up the potential slope asymptotically.
The condition (\ref{condition_finite-dilaton}) is rewritten as
\begin{equation}
\lambda < \gamma,
\label{condition_2}
\end{equation}
or
\begin{equation}
\lambda>\frac{D\gamma}{\sqrt{(D-4)^2-8(D-1)_2\c^2}}
~~~ \mbox{and} ~~~
 0<\gamma<\frac{D-4}{\sqrt{8(D-1)_2}}\;.
\label{condition_3}
\end{equation}
These regions are depicted in Fig.~\ref{parameter-region_2}.

\begin{figure}[th]
\hspace*{-2mm}
\includegraphics[width=3.8cm]{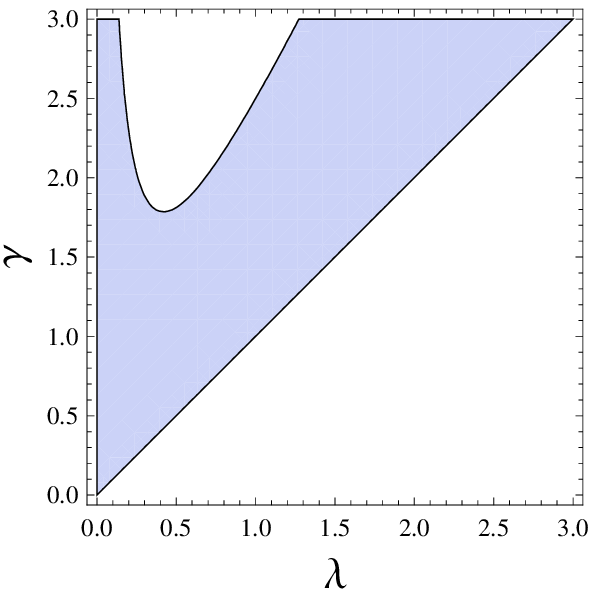}~
\includegraphics[width=3.8cm]{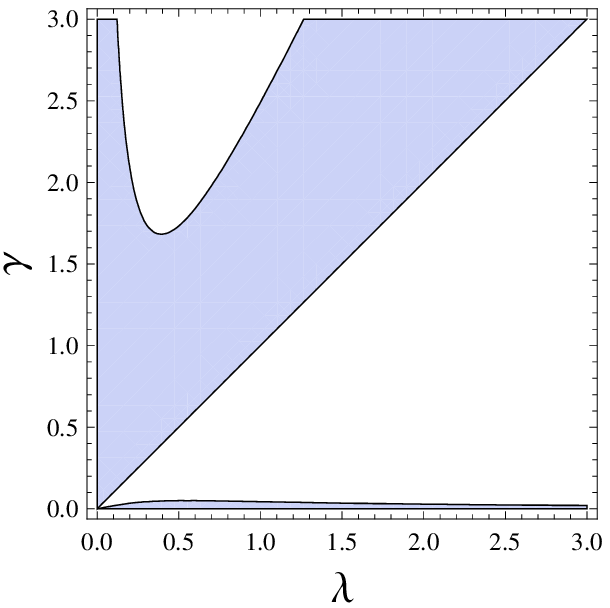}~
\includegraphics[width=3.8cm]{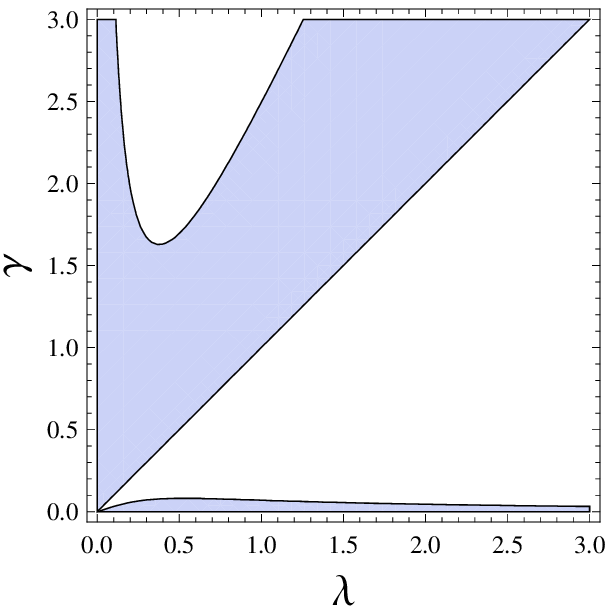}~
\includegraphics[width=3.8cm]{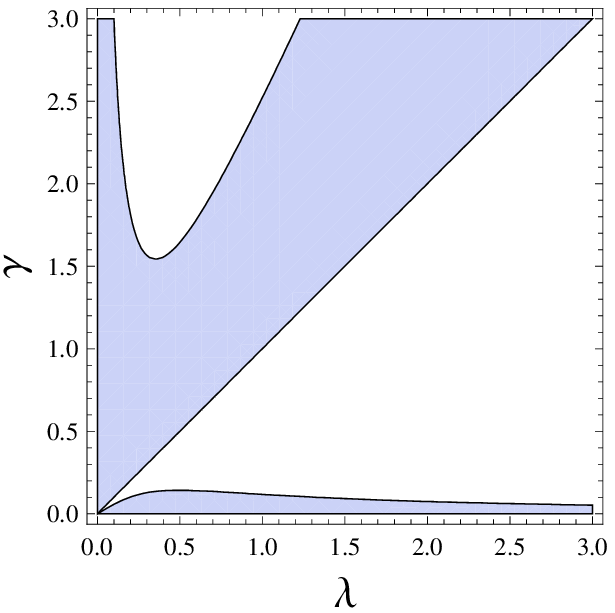}\\
\vspace{5mm}
\put(50,0){(a)}
\put(166,0){(b)}
\put(283,0){(c)}
\put(398,0){(d)}
\caption{The allowed regions from the conditions~\p{condition_1} and
\p{condition_finite-dilaton} for (a) $D=4$, (b) $D=5$, (c) $D=6$, and (d) $D=10$.
}
\label{para3}
\end{figure}

By superposing Figs.~\ref{parameter-region_1} and \ref{parameter-region_2},
we find that there are two separate allowed regions in the parameter
space~$(\la,\c)$. This is shown in Fig.~\ref{para3}.
There is a narrow region near $\gamma=0$ axis, but we did not find any relevant
solutions with parameters in this region when we integrate the basic equations
outwards from the event horizon numerically. Since these conditions are obtained
by the asymptotic behaviors of the field functions, it is expected that they
do not extend there because the spacetime hits singularity before
reaching the asymptotic region.


\section{Black hole solutions}

The basic equations (\ref{f1})--(\ref{f3}) do not have analytical solutions,
so we have to resort to the numerical method.
In the numerical analysis, we have to first choose the parameters for our black
hole solutions from the allowed regions of $\c$ and $\la$, and other parameters.
Considering the results in the previous section, we choose the following
parameters and conditions as a typical example in various dimensions:
\bea
\c = \frac12,~~
\la = \frac13,~~
\tilde\Lambda<0,~~
\phi_-=0,~~
\d_0=0,
\label{bc}
\ena
and expect that this choice gives the typical solutions.
In fact, it should not be difficult to get solutions for other choice if
it is in the allowed region.

We next fix the radius of the event horizon  $\tilde{r}_H$ and the cosmological
constant  $\tilde{\Lambda}$.
We then choose the value of the dilaton field $\phi_H$ at the horizon,
and determine the values of  other fields by \p{bhor}.
Among these, $\phi_H$ should be tuned such that $\phi_-=0$ in the asymptotic
behavior~\p{behavior}. Hence there is only one freedom of choosing $\tilde{r}_H$,
given a cosmological constant.

Once a solution for one $\tr_H$ and a fixed cosmological term is obtained, we can get
solutions for different $\tr_H$ but with the same $\tilde\Lambda$
by using the transformation~\p{sym1}. Under this transformation, we have
\bea
\tr &\to& \hat r \equiv a\tr,\nn
B(\tr) &\to& \hat{B}(\hat r) \equiv a^2 B(\tr), \nn
&&~~~~~~ = a^2 \Bigg[\tilde b_2 (\hat r/a)^2
- \frac{2 \tilde{M_{+}}}{(\hat r/a)^{\nu_{+}-2}}
- \frac{2\tilde{M}_0}{(\hat r/a)^{D-3}}\Bigg] \nn
&&~~~~~~ = \tilde b_2 \hat r^2
- \frac{2 a^{\nu_{+}} \tilde{M}_{+}}{\hat r^{\nu_{+}-2}}
- \frac{2 a^{D-1} \tilde{M}_0}{\hat r^{D-3}}.
\label{tr1}
\ena
Thus the gravitational mass scales like $a^{D-1} \tilde M_0$.
This means that the mass $\tilde M_0$ depends on the horizon radius as
\bea
\tilde M_0 \propto \tr_H^{\;D-1},
\label{gravmass1}
\ena
for a fixed cosmological constant.

Given a solution for a cosmological constant, we can generate solutions for
different cosmological constants but the same $\tr_H$ by using transformation \p{sym2}.
Suppose that we get a solution for a certain $\tilde\Lambda$ and a fixed $\tilde{r}_H$.
Using Eq.~\p{sym2}, we get a new solution by
\bea
B \to a^{2\c/(\c-\la)} B, ~~
\phi \to \phi + \frac{2}{\c-\la}\ln a,~~
\tilde \Lambda \to a^2 \tilde \Lambda.
\label{tr2}
\ena
Note that this does not shift $\phi_-$ in the asymptotic expansion~\p{behavior},
so the condition $\phi_-=0$ is not spoiled.
Under this transformation, the mass changes as
\bea
\tilde M_0 \to a^{2\c/(\c-\la)} \tilde M_0.
\ena
This means that when the cosmological constant is changed, the mass scales like
\bea
\tilde M_0 \propto |\tilde\Lambda|^{\c/(\c-\la)},
\label{gravmass2}
\ena
independently of our spacetime dimension.
When the condition~\p{condition_2} is satisfied, the power of $\tilde\Lambda$ is
positive. Thus the mass becomes larger as the magnitude of the cosmological constant
becomes larger. For our choice $\c=1/2$ and $\la=1/3$, this gives
$\tilde M_0 \sim |\tilde \Lambda|^3$.

As in our previous paper,~\cite{GOT} we present our solutions for $D=4,5,6$ and 10
because those in $D=7,8, 9$ are similar to the solution in $D=10$.

\subsection{Non-dilatonic case}
\label{non-dil}

It will be instructive to compare our results with the non-dilatonic case.
So let us derive some physical quantities for this case here. When the dilaton
field is absent (i.e., Einstein-Gauss-Bonnet system with cosmological constant),
we substitute $\phi\equiv 0$ and $\gamma=0$
into Eqs.~\p{f1} and \p{f2}, which can then be integrated to yield
\bea 
&&
\label{nd-B}
B(\tr) = \frac{1}{2(D-3)_4}
 \left(1 \mp \sqrt{1-\frac{4(D-3)_4}{\tilde{\ell}^2}
+ \frac{8(D-3)_4\bar{M}}{\tr^{D-1}}}\,
\right) \tr^{2},
\\
&&
\delta({\tr}) \equiv 0
\ena 
where $\bar{M}$ is an integration constant related to the conserved mass of
the black hole.~\cite{Deser&Tekin}
The solutions in minus branch have black hole horizon and approaches
the solutions in general relativity in the $\alpha_2 \to 0$ limit.
Hence this is the general relativity (GR) branch.
The solutions in the plus branch do not have a event horizon and the spacetime
is naked singular. This is called the GB branch.
The condition $B(\tr_H)=0$ gives the relation between $\tilde{M}$ and $\tr_H$ as
\bea
\bar M = \frac{1}{2\tilde{\ell}^2} \tr_H^{\;D-1}.
\label{gm}
\ena
This dependence is the same as Eq.~(\ref{gravmass1}).
In the pure vacuum (source-less) spacetime $\bar{M}=0$, we have
\bea 
B(\tr) = \frac{1}{2(D-3)_4}
 \left(1 \mp \sqrt{1-\frac{4(D-3)_4}{\tilde{\ell}^2} }\, \right) \tr^{2}
\equiv \frac{\tr^{2}}{\bar{\ell}_{\rm AdS}^2},
\ena 
where $\bar{\ell}_{\rm AdS}^2$ is square of the AdS curvature radius
in the non-dilatonic case given as
\bea 
\label{leff_nondil}
\bar{\ell}_{\rm AdS}^2 =
\frac{\bar{\ell}^2}{2}\left(1 \pm \sqrt{1-\frac{4(D-3)_4}{\bar{\ell}^2}}\:
\right) .
\ena 
It is then natural to define the new mass function $\tilde{\mu}$ (gravitational mass) by
\bea
B(\tr) = \frac{\tr^2}{\bar{\ell}_{\rm AdS}^2}- \frac{2 \tilde{\mu}(\tr)}{\tr^{D-3}} \,,
\ena 
By Eq.~(\ref{nd-B}),
\bea
\tilde{\mu}(\tr)= \pm\frac{1}{4(D-3)_4}
\Biggl[\sqrt{1-\frac{4(D-3)_4}{\tilde{\ell}^2} + \frac{8(D-3)_4\bar{M}}{\tr^{D-1}}}
-\sqrt{1-\frac{4(D-3)_4}{\tilde{\ell}^2}}\Biggr] \tr^{D-1},
\ena 
where the plus and minus signs are for the GR and the GB branches respectively.
The asymptotic value $\tilde{\mu}(\infty)$, which corresponds to the gravitational
mass, is related to $\bar M$ by
$\bar M = \pm \tilde\mu(\infty)\sqrt{1-{4(D-3)_4}/{\tilde{\ell}^2}}$.

\subsection{$D=4$ solution}

\begin{figure}[ht]
\includegraphics[width=7.6cm]{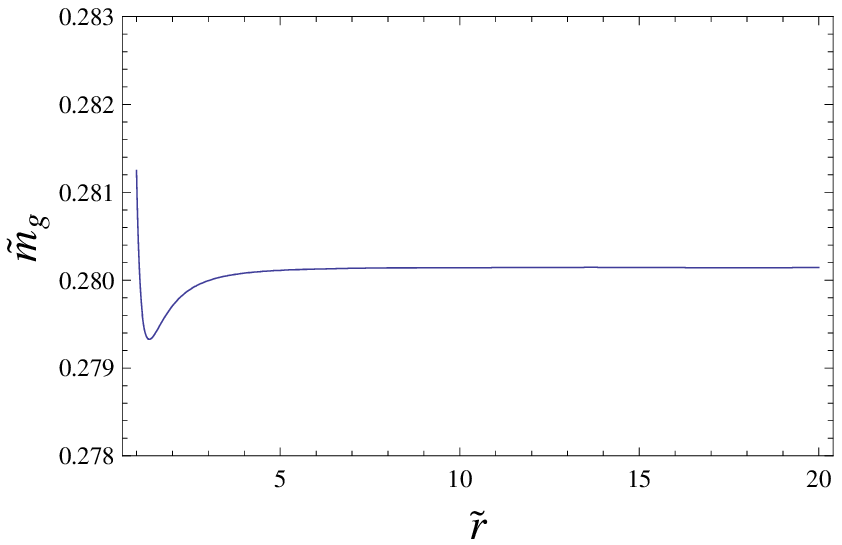}~~~~~~
\includegraphics[width=7.6cm]{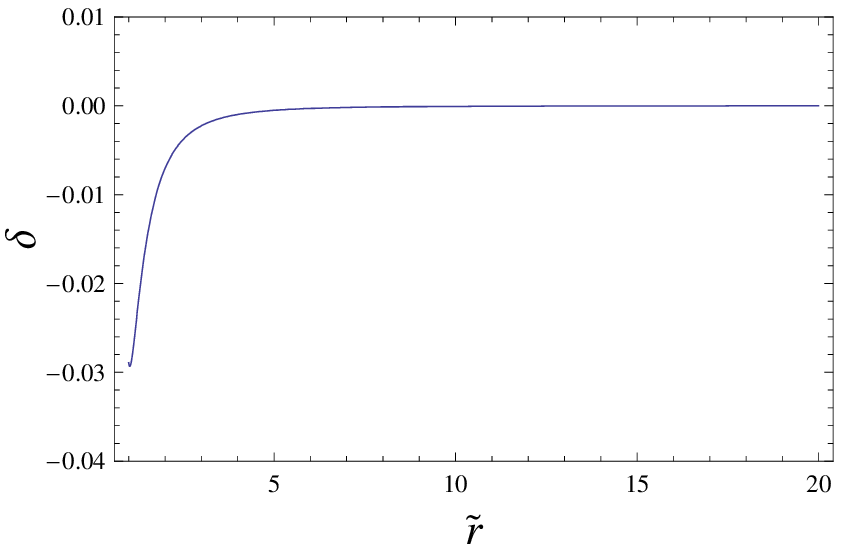}~~\\
\put(115,155){(a)}
\put(355,155){(b)}
\put(115,-15){(c)}
\includegraphics[width=7.6cm]{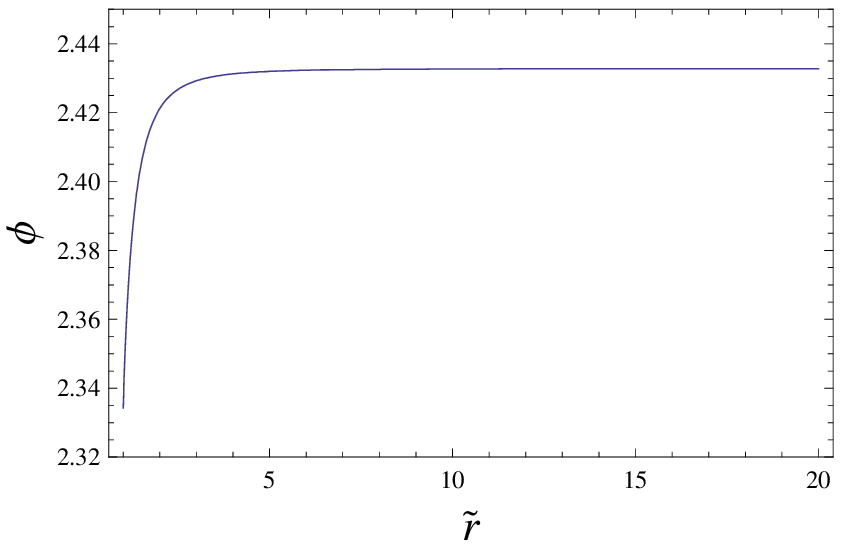}~~
\vs{5}
\caption{
The configurations of the field functions
(a) $\tm_g$, (b) $\d$ and (c) $\phi$ in four
dimensions for $\tr_H=1$ and $\tilde\Lambda=-3/2$.
}
\label{d4-dp}
\end{figure}

We first present the black hole solutions for $D=4$.
For the horizon radius $\tr_H=1$ and $\tilde \Lambda=-3/2$ $(\tilde{\ell}=2)$ with
the additional boundary conditions~\p{bhor} at the horizon, we integrate
the field equations from the event horizon to infinity. Then we find $\phi_H=2.33422$
in order to obtain $\phi_-=0$ as given in Eq.~\p{bc}, and
$\d_H=-0.02893$, $\phi_0=2.43279$ and $\tilde M_0=0.28014$.
The behaviors of $\tm_g$ (defined in Eq.~\p{mass}),
$\d$ and $\phi$ as functions of $\tr$ are depicted in Fig.~\ref{d4-dp}.
This solution corresponds to the dilatonic version of the minus branch solution
in the non-dilatonic case. We cannot find the counterpart of the plus branch
in non-dilatonic solution.
Solutions for other $\tr_H$ and cosmological constants is obtained from
this solution by the transformations~\p{tr1} and \p{tr2}, respectively.
It follows from Eqs.~(\ref{gravmass1}) and (\ref{gravmass2}) that
the gravitational mass $\tilde M_0$ is given by
\bea 
\tilde{M}_0 =0.28014 \biggl(\frac{2|\tilde\Lambda|}{3}\biggr)^{\! 3} \tr_H^{\;3} \, .
\ena 

\subsection{$D=5$ solution}

\begin{figure}[ht]
\includegraphics[width=7.6cm]{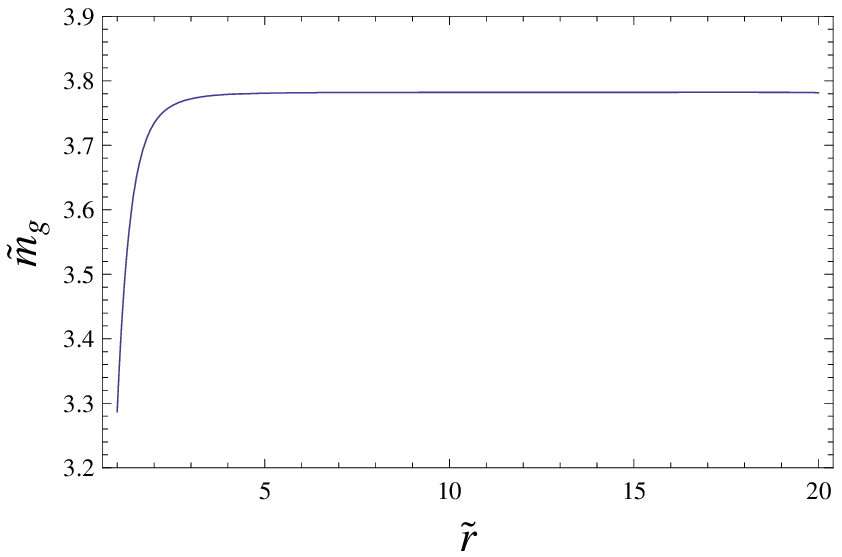}~~~~~~
\includegraphics[width=7.6cm]{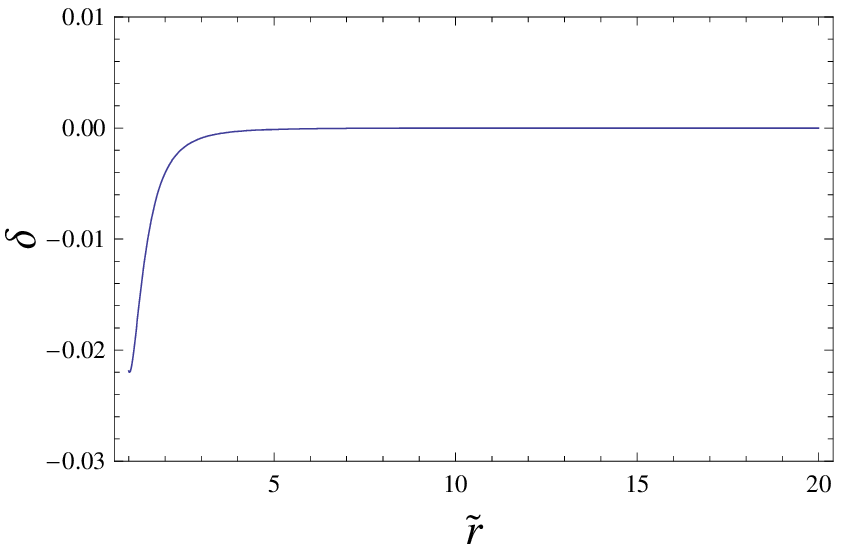}~~\\
\put(115,155){(a)}
\put(355,155){(b)}
\put(115,-15){(c)}
\includegraphics[width=7.6cm]{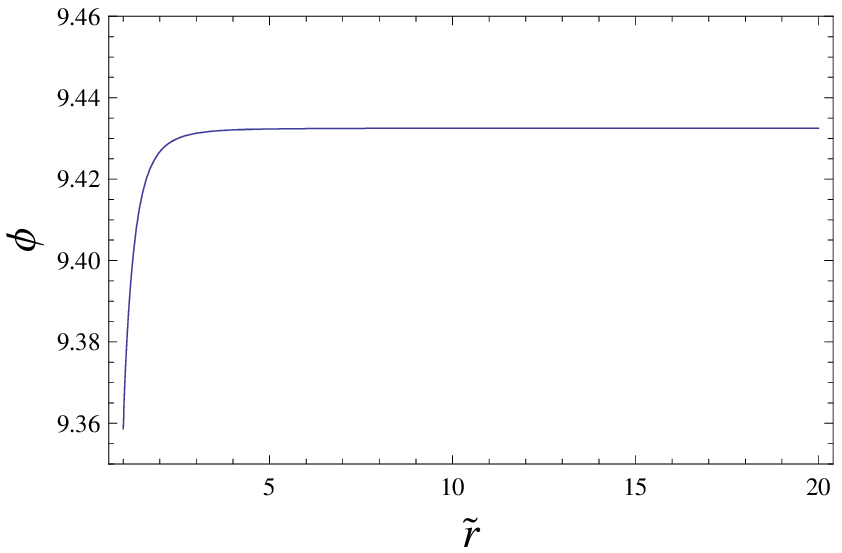}~~
\vs{5}
\caption{
The configurations of the field functions
(a) $\tm_g$, (b) $\d$ and (c) $\phi$ in five
dimensions for $\tr_H=1$ and $\tilde\Lambda=-3$.
}
\label{d5-dp}
\end{figure}

For the horizon radius $\tilde{r}_H=1$ and $\tilde\Lambda=
-3$ $(\tilde{\ell}=2)$ with the additional
boundary conditions~\p{bhor} at the horizon, we find
$\phi_H=9.35869$ in order to obtain $\phi_-=0$.
Then we find $\d_H=-0.02188, \phi_0=9.43249$ and $\tilde M_0=3.78189$.
The behaviors of $\tm, \d$ and $\phi$ as functions of $\tr$
are depicted in Fig.~\ref{d5-dp}.
Solutions for other $\tr_H$ and cosmological constants may be obtained from
this solution by the transformations~\p{tr1} and \p{tr2}, respectively.
The gravitational mass $\tilde M_0$ is given by the rules~\p{gravmass1}
and \p{gravmass2} as a function of the cosmological constant $\tilde\Lambda$
and the horizon radius $\tr_H$:
\bea 
\tilde{M}_0 =3.7819 \biggl(\frac{|\tilde\Lambda|}{3}\biggr)^{\! 3} \tr_H^{\;4} \, .
\ena 

\subsection{$D=6$ solution}

\begin{figure}[ht]
\includegraphics[width=7.6cm]{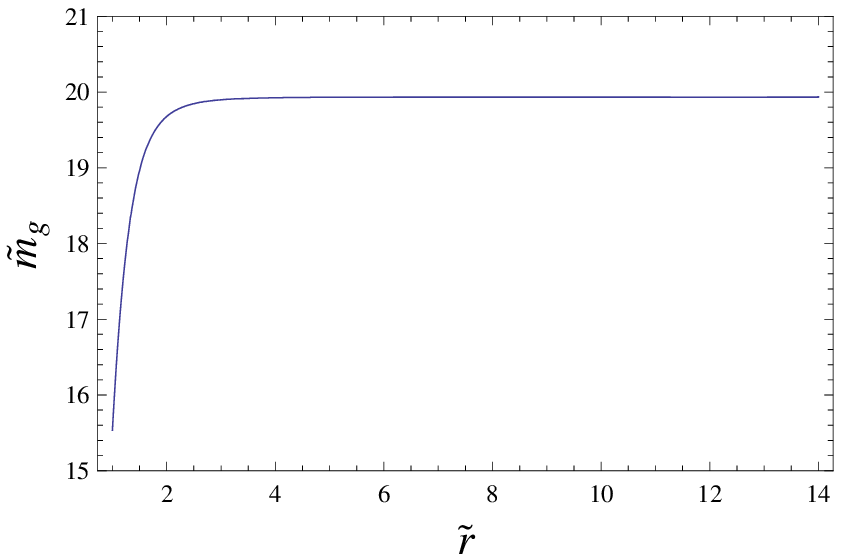}~~~~
\includegraphics[width=7.6cm]{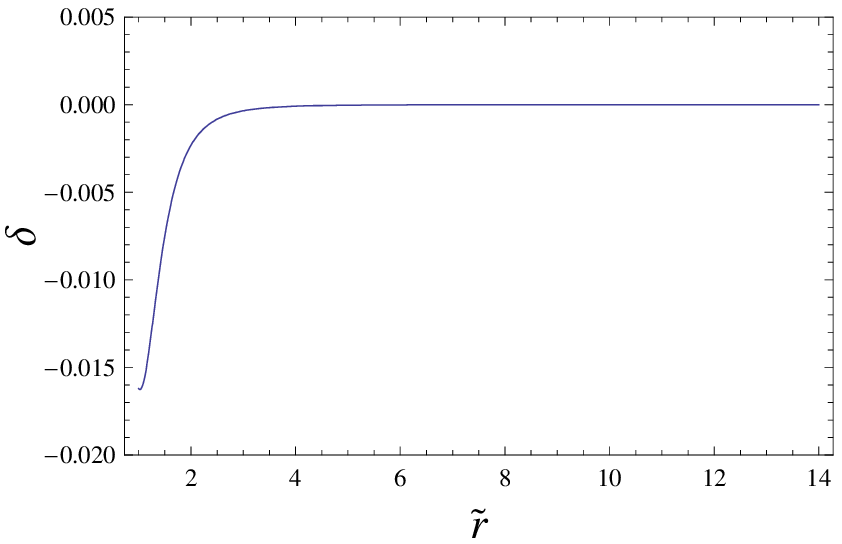}~~\\
\put(115,155){(a)}
\put(355,155){(b)}
\put(115,-15){(c)}
\includegraphics[width=7.6cm]{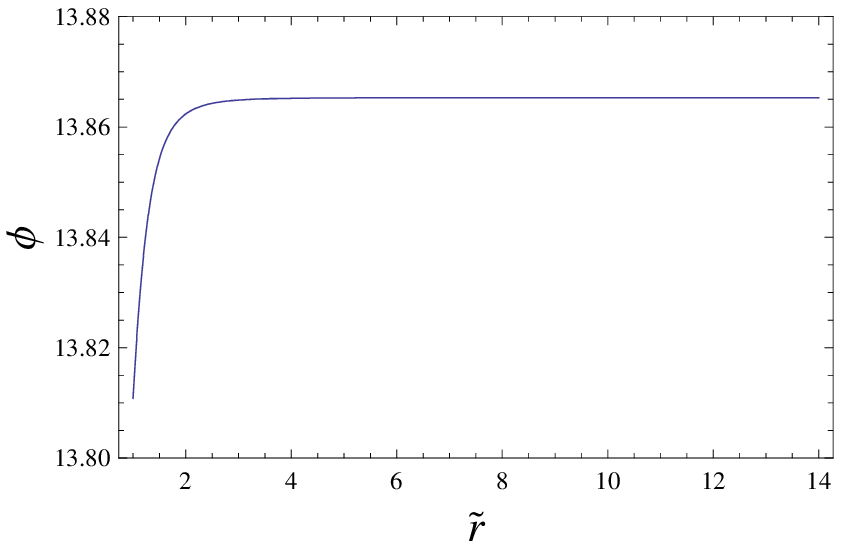}~~
\vs{5}
\caption{
The configurations of the field functions
(a) $\tm$, (b) $\d$ and (c) $\phi$  in six
dimensions for $\tr_H=1$ and $\tilde\Lambda=-5$.
}
\label{d6-dp}
\end{figure}

For the horizon radius $\tr_H=1$ and $\tilde\Lambda
=-5$ $(\tilde{\ell}=2)$ with the additional
boundary conditions~\p{bhor} at the horizon, we find
$\phi_H= 13.8108$ in order to obtain $\phi_-=0$.
Then we find $\d_H=-0.01621, \phi_0=13.86530$ and $\tilde M_0=19.93321$.
The behaviors of $\tm, \d$ and $\phi$ as functions of $\tr$,
which are depicted in Fig.~\ref{d6-dp}.
Solutions for other $\tr_H$ and cosmological constants may be obtained from
this solution by the transformations~\p{tr1} and \p{tr2}, respectively.
The gravitational mass $\tilde M_0$ is given by the rules~\p{gravmass1}
and \p{gravmass2} as a function of the cosmological constant $\tilde\Lambda$
and the horizon radius $\tr_H$:
\bea 
\tilde{M}_0 =19.933 \biggl(\frac{|\tilde\Lambda|}{5}\biggr)^{\! 3} \tr_H^{\;5} \, .
\ena 

\subsection{$D=10$ solution}

\begin{figure}[ht]
\includegraphics[width=7.6cm]{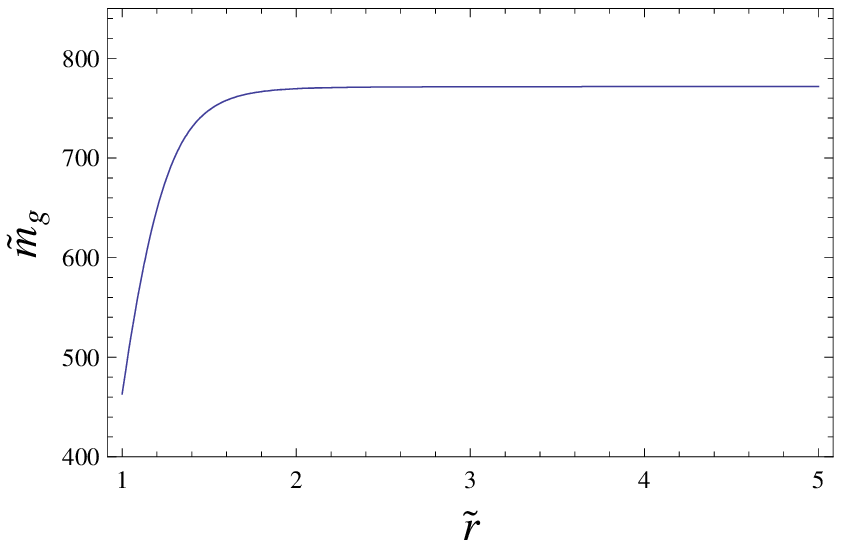}~~~~
\includegraphics[width=7.6cm]{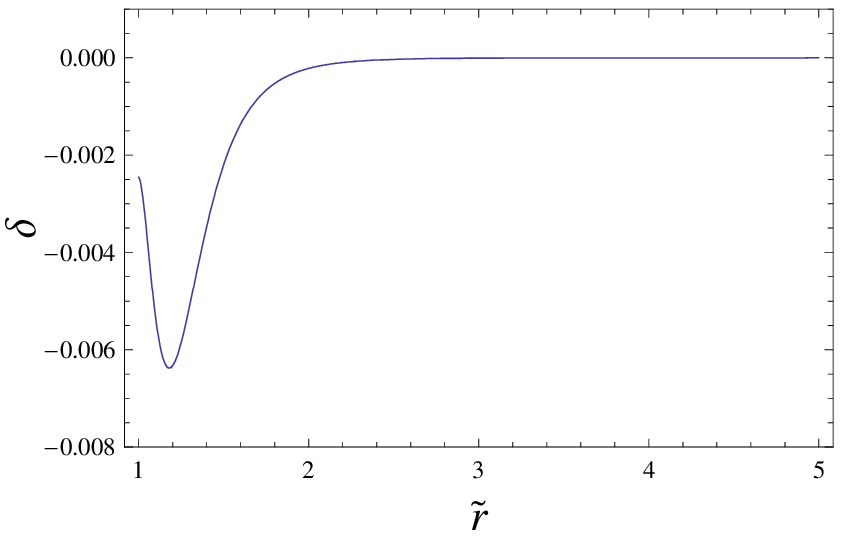}~~\\
\put(115,155){(a)}
\put(355,155){(b)}
\put(115,-15){(c)}
\includegraphics[width=7.6cm]{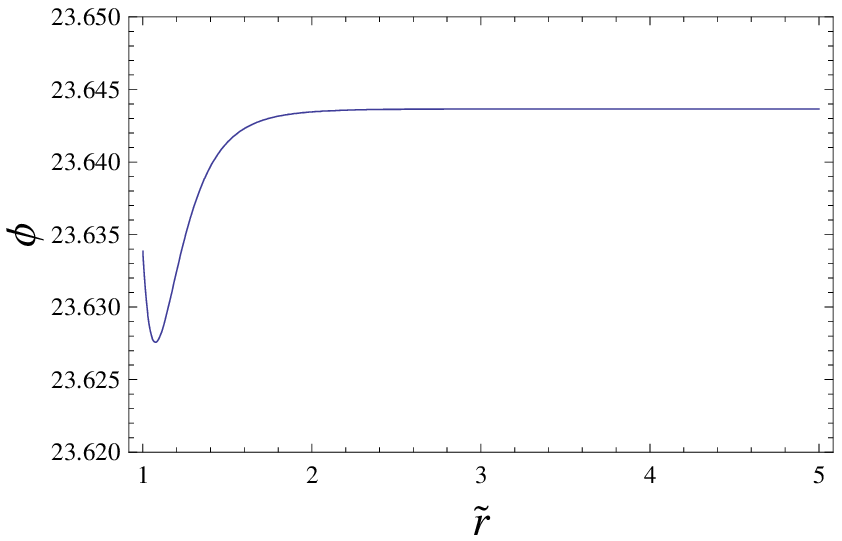}~~
\vs{5}
\caption{
The configurations of the field functions
(a) $\tm$, (b) $\d$ and (c) $\phi$  in ten
dimensions for $\tr_H=1$ and $\tilde\Lambda=-18$.
}
\label{d10-dp}
\end{figure}

For the horizon radius $\tr_H=1$ and $\tilde\Lambda
=-18$ $(\tilde{\ell}=2)$ with the additional
boundary conditions~\p{bhor} at the horizon, we find
$\phi_H= 23.6338$ in order to obtain $\phi_-=0$ as given in Eq.~\p{bc}.
Then we find $\d_H=-0.0024575, \phi_0=23.64366$ and $\tilde M_0=771.67622$.
The behaviors of $\tm, \d$ and $\phi$ as functions of $\tr$, are depicted
in Fig.~\ref{d10-dp}.
Solutions for other $\tr_H$ and cosmological constants may be obtained from
this solution by the transformations~\p{tr1} and \p{tr2}, respectively.
The gravitational mass $\tilde M_0$ is given by the rules~\p{gravmass1}
and \p{gravmass2} as a function of the cosmological constant $\tilde\Lambda$
and the horizon radius $\tr_H$:
\bea 
\tilde{M}_0 =771.68 \biggl(\frac{|\tilde\Lambda|}{18}\biggr)^{\! 3} \tr_H^{\;9} \, .
\ena 

\section{Conclusions and Discussions}
\label{CD}

We have studied the black hole solutions in dilatonic Einstein-GB theory with
the negative cosmological constant. The cosmological constant introduces
the Liouville type of potential for the dilaton field. We have taken the plane
symmetric spacetime, i.e., the $(D-2)$-dimensional hypersurface spanned by the
angular coordinates with vanishing constant curvature ($k=0$). The basic equations
have some symmetries which are used to generate the black hole solutions
with different horizon radius and the cosmological constant.

We have also examined the boundary conditions at the horizon and found that
there is no asymptotically AdS solution unless we introduce the cosmological
constant. By the asymptotic expansion at infinity, the power decaying rate of
the field variables are estimated. We have imposed the condition that the ``mass"
of the dilaton field satisfies the BF bound, which guarantees the stability of
the vacuum solution. By this condition, the values of the dilaton coupling
constant and the parameter of the Liouville potential are constrained.
For a typical choice of the parameters and boundary conditions, we were able
to construct AdS black hole solutions in various dimensions.

In the non-dilatonic case, there are two analytical solutions, one of which is
the black hole solutions (GR branch) and the other is the solution with the
naked singularity (GB branch). In the dilatonic case, we have chosen $\gamma=1/2$ and
$\lambda=1/3$ for the actual numerical analysis. The black hole solutions
are constructed in $D=4, 5, 6$ and $10$. We have checked that the dilaton field
climbs up its potential slope and takes constant values at infinity.
We have found that the relation of the gravitational mass and the horizon
radius of the black hole is
\bea 
\tilde{M}_0 \propto |\tilde\Lambda|^{\gamma/(\gamma-\lambda)} \tr_H^{\;D-1} \, .
\ena 

There are some remaining issues left for future works.
One of them is the thermodynamics of our black holes.
The Hawking temperature is given by the periodicity of the Euclidean time
on the horizon as
\bea 
\label{temp}
\tilde{T}_H ~&& =\frac{e^{-\d_H}}{4\pi}B_H'
\nonumber \\
&& =-\frac{e^{-\d_H}}{4(D-2)\pi}\tilde{\Lambda}\tilde{r}_H e^{\lambda \phi_H},
\ena 
where the first equality holds for any $k$ but we set $k=0$ in obtaining the second line.
It follows from the scaling symmetry (\ref{sym1}) that the temperature is proportional
to the horizon radius.
In the case of GB gravity, the entropy is not obtained by a quarter of the area
of the event horizon.
Along the definition of entropy in Ref.~\citen{Wald}, which originates from
the Noether charge associated with the diffeomorphism invariance of
the system, we obtain
\bea 
\tilde{S} = \frac{\tr_H^{D-2}\Sigma_{k}}{4}
\left[1+2(D-2)_3 \frac{ke^{-\c\phi_H}}{\tr_H^2} \right]-\tilde{S}_{min},
\ena 
where $S_{min}$ is added to make the entropy non-negative\cite{Clunan}.
(See also Ref.~\citen{CNO}.)
For the plane symmetric case $k=0$, we can set $S_{min}=0$, and the entropy is
proportional to $\tr_H^{\:D-2}$.
The $\tr_H$ dependence of the temperature and entropy shows that the thermodynamical
mass is estimated as $\tilde{M}_{thermo}\propto \tr_H^{\;D-1}$ by the first
law of black hole thermodynamics if we assume that the zero size black hole
has zero mass. This is same as that of the gravitational mass.

In this paper, we have assumed that $\tm^2<0$, which means that there is no growing
mode of the dilaton field asymptotically. However, even for the case with $\tm^2>0$,
the growing mode may be turned off by tuning the boundary value
$\phi_H$. The dilaton field of such solution is normalizable and decays faster
than that of the solutions presented in this paper. Furthermore,
the solution must be stable by the form of the non-tachyonic potential.
Hence it will be physically relevant. The search for this
solutions is left for future work.
On the other hand, even if the solution has a growing mode,
when the dilaton field diverges minus infinity and
$\tilde{b}_2 =0$, the system reduces to GR.
This situation is similar to Ref.~\citen{cai}.
The relevance of such solution should be investigated further.

It would be also interesting to extend our work to other spacetimes, including
$k=1$ and other topological black holes with the cosmological constant.
We plan to report our results on these cases in the future publication.

Finally we hope that our asymptotically AdS black hole solutions are
useful for examining properties of field theories via AdS/CFT correspondence.

\section*{Acknowledgements}

We would like to thank Christos Charmousis, Gary W. Gibbons, Hideo Kodama,
Kei-ichi Maeda, and Kengo Maeda for valuable discussions.
The work of Z.K.G. and N.O. was supported in part by the Grant-in-Aid for
Scientific Research Fund of the JSPS Nos. 20540283 and 06042.
The work of N.O. was also supported by the Japan-U.K. Research Cooperative Program.

\end{document}